\documentclass[prd,twocolumn,showpacs,floats,floatfix,nofootinbib]{revtex4}
\usepackage{graphicx}
\usepackage{dcolumn}
\usepackage{amssymb}
\usepackage{bm}
\def\spose#1{\hbox to 0pt{#1\hss}}

\def\lta{\mathrel{\spose{\lower 3pt\hbox{$\mathchar"218$}}
     \raise 2.0pt\hbox{$\mathchar"13C$}}}
\def\gta{\mathrel{\spose{\lower 3pt\hbox{$\mathchar"218$}}
     \raise 2.0pt\hbox{$\mathchar"13E$}}}
\newcommand{\be}{\begin{equation}}
\newcommand{\en}{\end{equation}}
\newcommand{\bea}{\begin{eqnarray}}
\newcommand{\ena}{\end{eqnarray}}
\newcommand{\ex}{\mbox{e}}
\newcommand{\dd}{\mbox{d}}

\def\setR{\mathbb{R}}
\def\setC{\mathbb{C}}

\newcommand{\ie}{\textsl{i.e.~}}

\newcommand{\eg}{\textsl{e.g.~}}
\newcommand{\etal}{\textsl{et al.~}}

\newcommand{\Hu}{{\cal H}} \newcommand{\Ka}{{\cal K}}
 
\newcommand{\GN}{G_{_\mathrm{N}}}

\newcommand{\lP}{\ell_{_\mathrm{Pl}}}

\begin{document}

\title{A non inflationary model with scale invariant cosmological
perturbations}

\author{Patrick Peter} \email{peter@iap.fr}
\affiliation{${\cal G}\setR\varepsilon\setC{\cal O}$ --
Institut d'Astrophysique de
 Paris, UMR7095 CNRS, Universit\'e Pierre \& Marie Curie, 98 bis
 boulevard Arago, 75014 Paris, France}

\author{Emanuel J. C. Pinho} \email{emanuel@cbpf.br}
\affiliation{Lafex - Centro Brasileiro de Pesquisas F\'{\i}sicas --
CBPF, \\ rua Xavier Sigaud, 150, Urca, CEP22290-180, Rio de Janeiro,
Brazil}

\author{Nelson Pinto-Neto} \email{nelsonpn@cbpf.br} \affiliation{Lafex
- Centro Brasileiro de Pesquisas F\'{\i}sicas -- CBPF, \\ rua Xavier
Sigaud, 150, Urca, CEP22290-180, Rio de Janeiro, Brazil}

\date{\today}

\begin{abstract}
We show that a contracting universe which bounces due to
quantum cosmological effects
and connects to the hot big-bang expansion phase, can produce
an almost scale invariant spectrum of perturbations provided the
perturbations are produced during an almost matter dominated era
in the contraction phase.
This is achieved using Bohmian solutions of the canonical Wheeler-de Witt
equation, thus treating both the background and the perturbations in
a fully quantum manner. We find a very slightly blue spectrum
($n_{_\mathrm{S}}-1>0$). Taking into account the spectral index
constraint as well as the CMB normalization measure yields an
equation of state that should be less than $\omega\lesssim 8\times
10^{-4}$, implying $n_{_\mathrm{S}}-1 \sim
\mathcal{O}\left(10^{-4}\right)$, and that the characteristic size
of the Universe at the bounce is $L_0 \sim 10^3 \lP$, a region where
one expects that the Wheeler-DeWitt equation should be valid without 
being spoiled by string or loop quantum gravity effects.
\end{abstract}

\maketitle

\section{Introduction}

When the theory of cosmological perturbations~\cite{MFB} is applied to
a background cosmological model and their initial spectra can be
justified in physical terms, as it is the case of inflationary and
nonsingular models where the horizon problem is solved, one obtains
definite predictions concerning the spectrum of primordial scalar and
tensor perturbations in this background, establishing the initial
conditions for structure formation and the angular power spectrum of
the anisotropies of the cosmic microwave background radiation
(CMBR). Since the release of the first data obtained from observations
of these anisotropies in 1992~\cite{cobe}, primordial cosmological
models have been confronted with new observational facts~\cite{WMAP},
bringing them definitely out of the arena of speculations based almost
uniquely on theoretical and \ae{}sthetical arguments. In fact, many
interesting cosmological backgrounds have been falsified since
then~\cite{falsify}, while the inflationary paradigm~\cite{inflation}
has, on the contrary, been widely confirmed and is now accepted as
part of the ``standard'' model of cosmology.

Some of the primordial cosmological backgrounds proposed in
the literature are quantum
cosmological models which share the attractive properties of
exhibiting neither singularities nor
horizons~\cite{pinto,pinto2,fabris}, leading the universe evolution
through a bouncing phase due to quantum effects, and a contracting
phase before the bounce. They constitute an example of a bouncing
model, a possibility that has attracted the attention of many
authors~\cite{bounce}, without the presence of any phantom field.
These features of the background introduce a new picture for the
evolution of cosmological perturbations: vacuum initial conditions may
now be imposed when the Universe was very big and almost flat, and
effects due to the contracting and bouncing phases, which are not
present in the standard background cosmological model, may change the
subsequent evolution of perturbations in the expanding phase. Hence,
it is quite important to study the evolution of perturbations in these
quantum backgrounds to confront them with the data.

The present paper is the fourth of a series~\cite{tens1,tens2,scalar}
where the theory of cosmological perturbations is obtained and
simplified without assuming any dynamics satisfied by the
background. This is a necessary prerequisite if one wants to study the
propagation of perturbations on a quantized background. The usual
theory of cosmological perturbations with their simple
equations~\cite{MFB} relies essentially on the assumptions that the
background is described by pure classical General Relativity (GR),
while the perturbations thereof stem from quantum fluctuations. It is
a semiclassical approach, where the background is classical and the
perturbations are quantized. A full quantum treatment of both
background and perturbations has already been constructed in
Ref.~\cite{halli}, but rather complicated equations were obtained,
even at first order, due to the fact that the background does not
satisfy classical Einstein's equations. In Refs.~\cite{tens1,scalar},
we have managed to put these complicated equations in simple forms,
very similar to the equations obtained in Ref.~\cite{MFB}, through the
implementation of canonical transformations and redefinitions of the
lapse function without ever using the background classical equations.
These expressions happen to become identical to those of ~\cite{MFB}
when the background behaves as a pre-determined function of time,
which is perfectly consistent with the idea of quantization if we work
with an ontological interpretation of quantum mechanics~\cite{hol}. In
Ref.~\cite{tens2}, these results have been applied to obtain the
possible power spectra of tensor perturbations in different quantum
cosmological models using an hydrodynamical description. The aim of
this paper is to apply the results of Ref.~\cite{scalar} to obtain the
power spectra of scalar perturbations in such quantum models and
confront the results with observational data~\cite{WMAP}.

The paper is organized as follows. In Sec.~\ref{sec:quant} we
summarize the results of Refs.~\cite{scalar} obtaining the simple
equations which govern the dynamics of quantum perturbations in the
quantum backgrounds of Refs.~\cite{pinto,fabris}. In
Sec.~\ref{sec:mixe} we obtain the spectral index for long
wavelengths of scalar perturbations in these quantum backgrounds,
and in Sec.~\ref{sec:nume} we confirm these results numerically,
also obtaining their amplitude and constraining the free parameters
of the theory with observational data. We end in Sec.~V with conclusions
and discussions.

\section{Quantization of the background and perturbations}
\label{sec:quant}

The action we shall begin with is that of General Relativity (GR) with
a perfect fluid, the latter being described as in
Ref.~\cite{MFB}\footnote{One can also use the formalism due to
Schutz~\cite{Schutz}, obtaining the same results.}, \ie
\begin{equation}
\mathcal{S}= \mathcal{S}_{_\mathrm{GR}} + \mathcal{S}_\mathrm{fluid}
= -\frac{1}{6\lP^2} \int \sqrt{-g} R \dd^4 x - \int \sqrt{-g}
\epsilon \dd^4 x, \label{action}
\end{equation}
where $\lP=(8\pi\GN/3)^{1/2}$ is the Planck length in natural units
($\hbar=c=1$), $\epsilon$ is the perfect fluid energy density whose
pressure $p$ is provided by the relation $p=\omega\epsilon$,
$\omega$ being a nonvanishing constant.

Let the geometry of spacetime be given by
\begin{equation}
\label{perturb}
g_{\mu\nu}=g^{(0)}_{\mu\nu}+h_{\mu\nu},
\end{equation}
where $g^{(0)}_{\mu\nu}$ represents a homogeneous and isotropic
cosmological background,
\begin{equation}
\label{linha-fried}
\dd s^{2}=g^{(0)}_{\mu\nu}\dd x^{\mu}\dd x^{\nu}=N^{2}(t)\dd t^2 -
a^{2}(t)\delta_{ij}\dd x^{i}\dd x^{j},
\end{equation}
where we are restricted to a flat spatial metric, and the $h_{\mu\nu}$
represents linear scalar perturbations around it which we decompose
into
\begin{eqnarray}
\label{perturb-componentes}
h_{00}&=&2N^{2}\phi \nonumber \\
h_{0i}&=&-NaB_{,i} \\
h_{ij}&=&2a^{2}(\psi\gamma_{ij}-E_{,ij}). \nonumber
\end{eqnarray}

Substituting Eqs.~(\ref{perturb-componentes}) and
(\ref{linha-fried}) into the Einstein-Hilbert action (\ref{action}),
performing Legendre and canonical transformations, redefining $N$
with terms which do not alter the equations of motion up to first
order, all this without ever using the background equations of
motion, the Hamiltonian up to second order is simplified to (see
Ref.~\cite{scalar} for details)
\begin{equation}
\label{hfinal-vinculos-escalares-hidro} H=N\left[ H_0^{(0)} +
H_0^{(2)}\right] +\Lambda_N P_N + \int \dd^{3}x\phi\pi_{\psi}+ \int
\dd^{3}x\Lambda_{\phi}\pi_{\phi},
\end{equation}
where
\begin{equation}
\label{h00} H_0^{(0)}\equiv
-\frac{l^{2}P_{a}^{2}}{4aV}+\frac{P_{T}}{a^{3\omega}},
\end{equation}
and
\begin{equation}
\label{h02} H_0^{(2)}\equiv \frac{1}{2a^{3}}\int
\dd^{3}x\pi^{2}+\frac{a\omega}{2} \int \dd^{3}x v^{,i}v_{,i}.
\end{equation}
The quantities $N$, $\phi$, $\Lambda_N$ and $\Lambda_{\phi}$ play the
role of Lagrange multipliers of the constraints $H_0^{(0)} +
H_0^{(2)}\approx 0$, $\pi_{\psi}\approx 0$, $P_{N}\approx0$, and
$\pi_{\phi}\approx 0$, respectively. The momenta $P_a$, $\pi_{\phi}$,
$\pi_{\psi}$, $P_N$ and $P_T$ are conjugate respectively to $a, \phi,
\psi, N, T$, this last variable playing the role of time.

The variable $v$ is related with the gauge invariant Bardeen
potential $\Phi$ (see Ref.~\cite{scalar}) through
\begin{equation}
\label{vinculo-simples}
\Phi^{,i}\,_{,i} =
-\frac{3\lP^2\sqrt{(\omega+1)\epsilon_0}}{2\sqrt{\omega}}a
\biggl(\frac{v}{a}\biggr)' ,
\end{equation}
which coincides with equation (12.8) of Ref.~\cite{MFB} relating $v$
and $\Phi$ \be \Phi^{,i}\,_{,i} = -\sqrt{\frac{3}{2}}
\frac{\lP\Hu\gamma}{c_\mathrm{s}^2} \left(
\frac{v}{z}\right)',
\label{vPhi} \en where
\be \gamma \equiv 1-\frac{\Hu'}{\Hu^2},\ \ \ \
z\equiv\frac{a}{c_\mathrm{s}} \sqrt{\gamma},\en and $c_\mathrm{s}$
($c^2_\mathrm{s} \equiv \dd p_0/\dd \epsilon_0 = p_0'/\epsilon_0'$)
is the velocity at which density perturbations propagates, and when
the classical equations of motion, that can be recast in the form
$$
\epsilon_0+p_0 = \epsilon_0 (1+\omega) = \frac{2\Hu^2\gamma}{3\lP^2
  a^2},
$$
are used.

The above quantities have been redefined in order to be
dimensionless. For instance, the physical scale factor $a_{\rm
phys}$ can be obtained from the dimensionless $a$ present
in~(\ref{h00}) through $a_{\rm phys}=\lP a/\sqrt{V}$, where $V$ is
the comoving volume of the background spacelike hypersurfaces,
which we suppose to be compact. The
constraint $H_0^{(0)} + H_0^{(2)}$ is the one which generates the
dynamics, yielding the correct Einstein equations both at zeroth and
first order in the perturbations, as can be checked explicitly.  The
others imply that $N$, $\phi$, and $\psi$ are not relevant. The
unique perturbed degree of freedom is $v$, as obtained in
Ref.~\cite{MFB}. We would like to emphasize again that in order to
obtain the above results, no assumption has been made about the
background dynamics:
Hamiltonian~(\ref{hfinal-vinculos-escalares-hidro}) is ready to be
applied in the quantization procedure.

In the Dirac quantization procedure, the first class constraints must
annihilate the wave functional $\Psi\left[
N,a,\phi(x^i),\psi(x^i),v(x^i),T\right]$, yielding
\begin{equation}
\label{vinculos-quanticos}
\frac{\partial}{\partial N} \Psi = \frac{\delta}{\delta\phi} \Psi =
\frac{\delta}{\delta\psi} \Psi = H\Psi =0.
\end{equation}
The first three equations impose that the wave functional does not
depend on $N$, $\phi$ and $\psi$: as mentioned above, $N$ and
$\phi$ are, respectively, the homogeneous and inhomogeneous parts
of the total lapse function, which are just Lagrange multipliers
of constraints, and $\psi$ has been substituted by $v(x^i)$, the
unique degree of freedom of scalar perturbations, as expected.

\begin{figure*}[t]
\vskip5mm\hskip-5mm
\includegraphics[width=8.0cm]{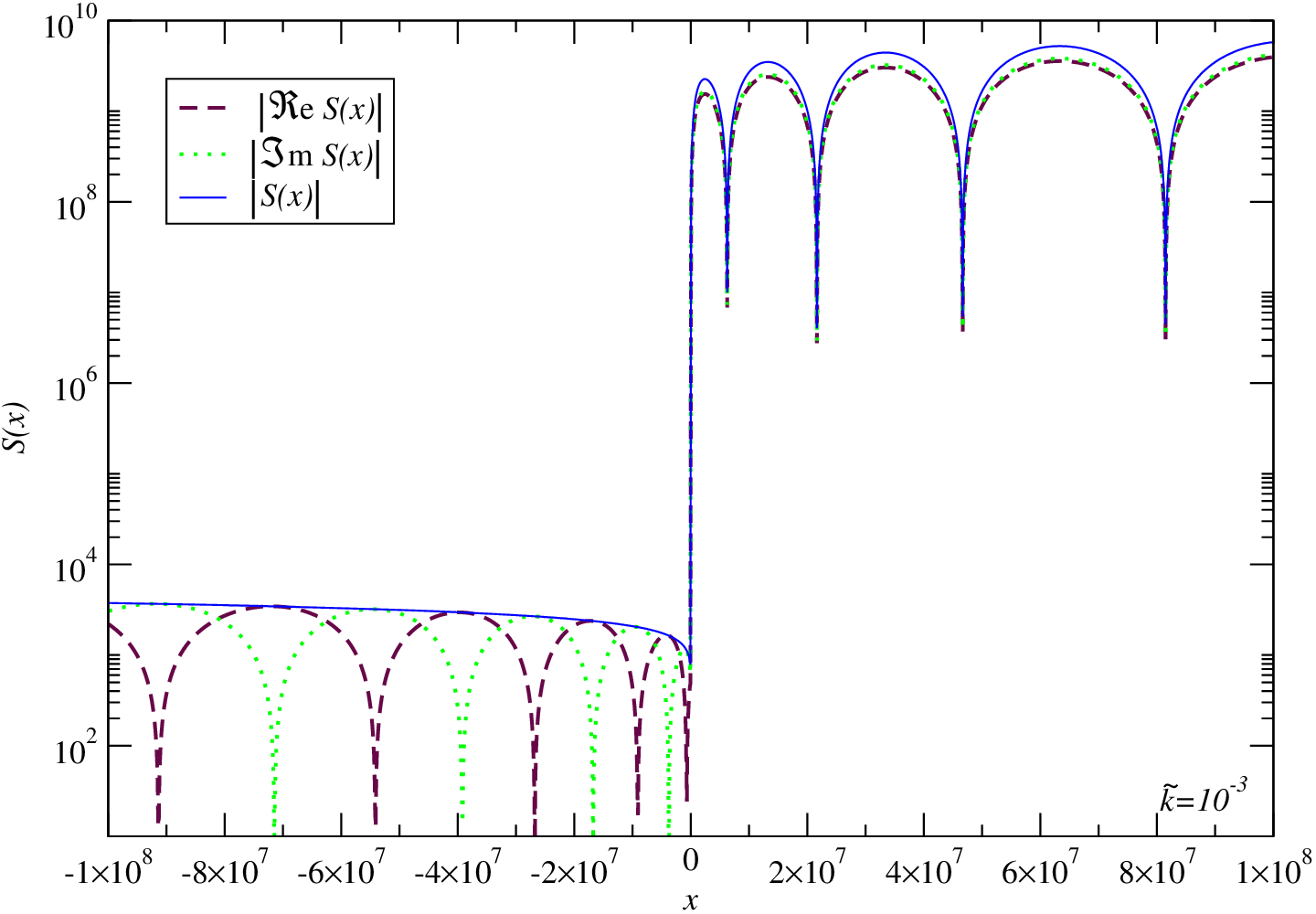}
\hskip5mm
\includegraphics[width=8.0cm]{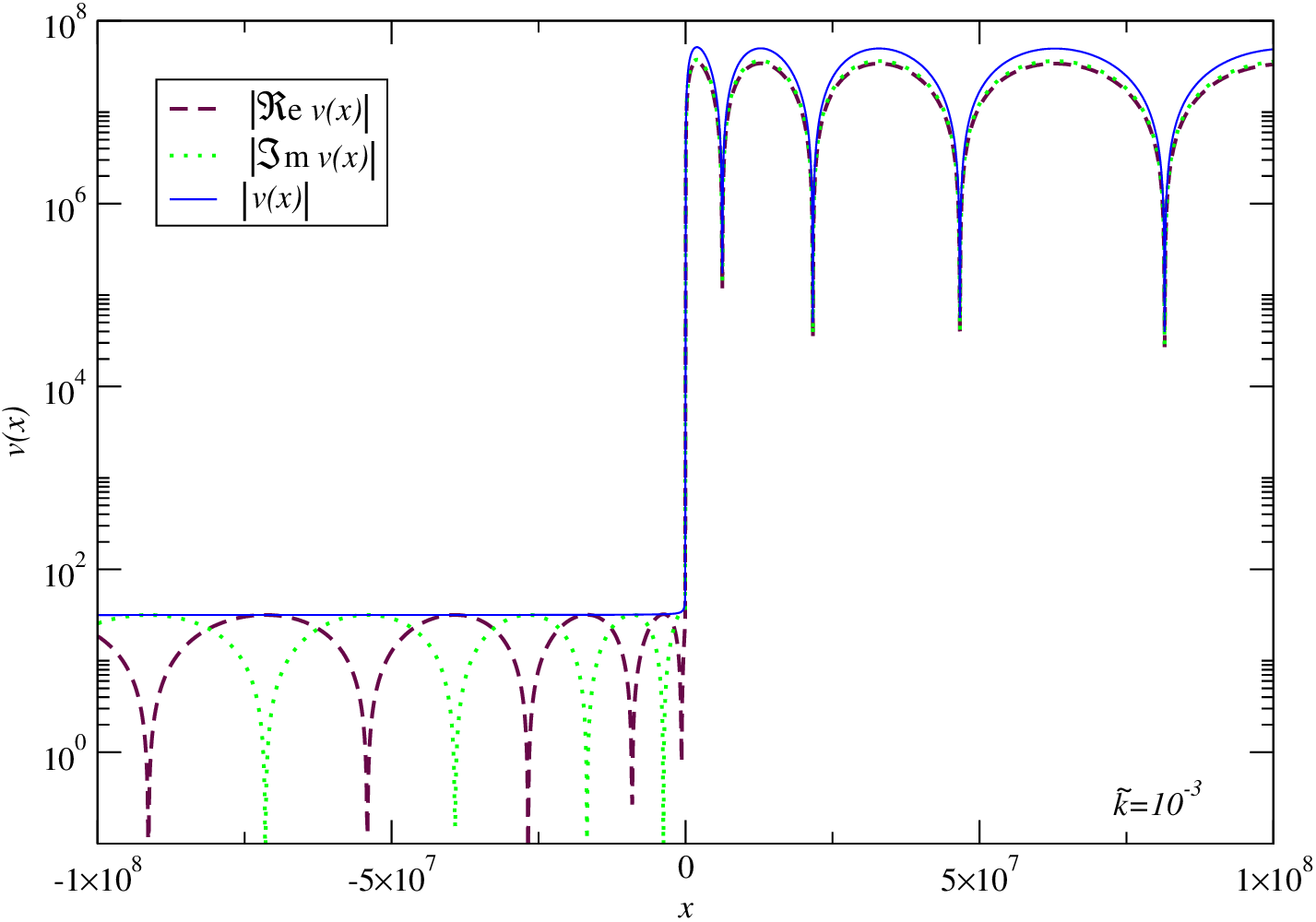}
\caption{Time evolution of the scalar mode function for the equation
of state $\omega=0.1$ in the one-fluid model of the bounce. The left
panel shows the full time evolution which was computed, i.e., the
function $S(x)$, while the right panel shows $v(x)$ itself, both
plots having $\tilde k=10^{-3}$. For $x<0$, there are oscillations
only in the real and imaginary parts of the mode, so the amplitude
shown is a non oscillating function of time. It however acquires an
oscillating piece after the bounce has taken place.} \label{fig:v01}
\end{figure*}

As $P_T$ appears linearly in $H$, and making the gauge choice
$N=a^{3\omega}$, one can interpret the $T$ variable as a time
parameter. Hence, the equation
\begin{equation}
\label{schroedinger} H\Psi=0
\end{equation}
assumes the Schr\"odinger form
\begin{eqnarray}
 i \frac{\partial}{\partial T}\Psi &=&\frac{1}{4} \left\{
a^{(3\omega-1)/2}\frac{\partial}{\partial a} \left[
a^{(3\omega-1)/2}\frac{\partial}{\partial a}\right]
\right\}\Psi\nonumber\\&-&\frac{a^{3\omega-1}}{2}\int
\dd^3x\frac{\delta^2}{\delta
v^2}\Psi+\frac{a^{3\omega+1}\omega}{2}\int \dd^3x v^{,i}v_{,i}\Psi
,\cr
& &
\end{eqnarray}
where we have chosen the factor ordering in $a$ in order to yield a
covariant Schr\"odinger equation under field redefinitions.

If one makes the ansatz
\begin{equation}
\label{ansatz} \Psi[a,v,T]=\Psi_{(0)}(a,T)\Psi_{(2)}[a,v,T],
\end{equation}
where $\Psi_{(0)}(a,T)$ satisfies the equation
\begin{equation}
\label{schroedinger-separado-fundo}  i  \frac{\partial}{\partial
T} \Psi_{(0)}(a,T)=\frac{a^{(3\omega-1)/2}}{4}
\frac{\partial}{\partial a} \left[
a^{(3\omega-1)/2}\frac{\partial}{\partial a}\right] \Psi_{(0)}(a,T),
\end{equation}
then we obtain for $\Psi_{(2)}[a,v,T]$ the equation.
\begin{eqnarray}
\label{schroedinger-separado-perturb}  i \frac{\partial}{\partial T}
\Psi_{(2)}(a,v,T)&=&-\frac{a^{(3\omega-1)}}{2}\int
\dd^3x\frac{\delta^2}{\delta v^2}\Psi_{(2)}(a,v,T)
\nonumber \\
&+&\frac{\omega a^{(3\omega+1)}}{2}\int \dd^3x
v^{,i}v_{,i}\Psi_{(2)}(a,v,T).\nonumber\\
\end{eqnarray}
Terms involving $\Psi_{(2)}(a,v,T)$ can be neglected in
Eq.~(\ref{schroedinger-separado-fundo}), either through a judicious
choice of the $a$ dependence of $\Psi_{(2)}$ \cite{tens2}, or because
quantum perturbations initiated in a vacuum quantum state should not
contribute to it \cite{halli}. In another perspective, using the
Bohmian approach \cite{hol}, one can write
$\Psi[a,v,T]=\Psi_{(0)}(a,T)\Psi_{(2)}[v,T]$ and consider $a(T)$
appearing in the equation for $\Psi_{(2)}[v,T]$ as a prescribed
function of time, the quantum Bohmian trajectory, obtained from the
zeroth order equation for $\Psi_{(0)}(a,T)$.

Going on with the ontological Bohm-de Broglie interpretation of
quantum mechanics, where quantum trajectories can be defined,
Eq.~(\ref{schroedinger-separado-perturb}) can be further simplified if
one uses Eq.~ (\ref{schroedinger-separado-fundo}) to obtain background
quantum Bohmian trajectories $a(T)$ as in
Refs.~\cite{pinto,pinto2,fabris}. This can be done as follows: we
change variables to
$$\chi=\frac{2}{3} (1-\omega)^{-1} a^{3(1-\omega)/2},$$ obtaining the
simple equation
\begin{equation}
i\frac{\partial\Psi_{(0)}(a,T)}{\partial T}= \frac{1}{4}
\frac{\partial^2\Psi_{(0)}(a,T)}{\partial \chi^2}. \label{es202}
\end{equation}
This is just the time reversed Schr\"odinger equation for a one
dimensional free particle constrained to the positive axis. As $a$
and $\chi$ are positive, solutions which have unitary evolution must
satisfy the condition
\begin{equation}
\label{cond27} \Psi_{(0)}^{\star}\frac{\partial\Psi_{(0)}}{\partial
\chi} -\Psi_{(0)}\frac{\partial\Psi_{(0)}^{\star}}{\partial
  \chi}\Biggl|_{\chi=0}=0
\end{equation}
(see Ref.~\cite{fabris} for details). We will choose the initial
normalized wave function
\begin{equation}
\label{initial}
\Psi_{(0)}^{(\mathrm{init})}(\chi)=\biggl(\frac{8}{T_0\pi}\biggr)^{1/4}
\exp\left(-\frac{\chi^2}{T_0}\right) ,
\end{equation}
where $T_0$ is an arbitrary constant. The Gaussian
$\Psi_{(0)}^{(\mathrm{init})}$ satisfies condition (\ref{cond27}).

Using the propagator procedure of Refs.~\cite{pinto,fabris}, we
obtain the wave solution for all times in terms of $a$:
\begin{widetext}
\begin{equation}\label{psi1t}
\Psi_{(0)}(a,T)=\left[\frac{8 T_0}{\pi\left(T^2+T_0^2\right)}
\right]^{1/4}
\exp\biggl[\frac{-4T_0a^{3(1-\omega)}}{9(T^2+T_0^2)(1-\omega)^2}\biggr]
\exp\left\{-i\left[\frac{4Ta^{3(1-\omega)}}{9(T^2+T_0^2)(1-\omega)^2}
+\frac{1}{2}\arctan\biggl(\frac{T_0}{T}\biggr)-\frac{\pi}{4}\right]\right\}
.
\end{equation}
\end{widetext}

Due to the chosen factor ordering, the probability density
$\rho(a,T)$ has a non trivial measure and it is given by
$\rho(a,T)=a^{(1-3\omega)/2}\left|\Psi_{(0)}(a,T)\right|^2$.  Its
continuity equation coming from Eq.~(\ref{es202}) reads
\begin{equation}
\label{cont} \frac{\partial\rho}{\partial T}
-\frac{\partial}{\partial a}\biggl[\frac{a^{(3\omega-1)}}{2}
\frac{\partial S}{\partial a}\rho\biggr]=0 ,
\end{equation}
which implies in the Bohm interpretation that
\begin{equation}
\label{guidance} \dot{a}=-\frac{a^{(3\omega-1)}}{2} \frac{\partial
S}{\partial a} ,
\end{equation}
in accordance with the classical relations $\dot{a}=\{a,H\}=
-a^{(3\omega-1)}P_a/2$ and $P_a=\partial S/\partial a$.

Inserting the phase of (\ref{psi1t}) into Eq.~(\ref{guidance}), we
obtain the Bohmian quantum trajectory for the scale factor:
\begin{equation}
\label{at} a(T) = a_0
\left[1+\left(\frac{T}{T_0}\right)^2\right]^\frac{1}{3(1-\omega)} .
\end{equation}
Note that this solution has no singularities and tends to the
classical solution when $T\rightarrow\pm\infty$. Remember that we
are in the gauge $N=a^{3\omega}$, and $T$ is related to conformal
time through
\begin{equation}
\label{jauge} N\dd T = a \dd \eta \quad \Longrightarrow \dd\eta =
\left[a(T)\right]^{3\omega-1} \dd T.
\end{equation}
The solution (\ref{at}) can be obtained for other initial wave
functions (see Ref.~\cite{fabris}).

The Bohmian quantum trajectory $a(T)$ can be used in
Eq.~(\ref{schroedinger-separado-perturb}). Indeed, since one can
view $a(T)$ as a function of $T$, it is possible to implement the
time dependent canonical transformation generated by
\begin{eqnarray}
\label{qct0} U &=& \exp\left[ i \left( \int \dd^3 x \frac{\dot{a}
v^2}{2a} \right) \right] \times\\
& & \times \exp\left\{ i \left[ \int \dd^3 x \left(
\frac{v\pi + \pi v}{2} \right) \ln\left( \frac{1}{a} \right)
\right]\right\}.
\end{eqnarray}
As $a(T)$ is a given quantum trajectory coming from Eq.~(\ref{es202}),
Eq.~(\ref{qct0}) must be viewed as the generator of a time dependent
canonical transformation to Eq.~(\ref{es202}). It yields, in terms of
conformal time, the equation for $\Psi_{(2)}[v,a(\eta),\eta]$
\begin{equation}
i\frac{\partial\Psi_{(2)}}{\partial \eta}= \int \dd^3 x
\left(-\frac{1}{2} \frac{\delta^2}{\delta v^2} +
\frac{\omega}{2}v_{,i} v^{,i} - \frac{{a''}}{2a}v^2 \right)
\Psi_{(2)}. \label{NENSE}
\end{equation}

This is the most simple form of the Schr\"odinger equation which
governs scalar perturbations in a quantum minisuperspace model with
fluid matter source. The corresponding time evolution equation for
the operator $v$ in the Heisenberg picture is given by
\begin{equation}
v''-\omega v^{,i}_{\ ,i}-\frac{a''}{a}v=0,
\end{equation}
where a prime means derivative with respect to conformal time. In
terms of the normal modes $v_k$, the above equation reads
\begin{equation}
\label{equacoes-mukhanov} v''_k+\biggl(\omega
k^2-\frac{{a''}}{a}\biggr)v_k=0.
\end{equation}
These equations have the same form as the equations for scalar
perturbations obtained in Ref.~\cite{MFB}. This is quite natural since
for a single fluid with constant equation of state $\omega$, the pump
function $z''/z$ obtained in Ref.~\cite{MFB} is exactly equal to
$a''/a$ obtained here. The difference is that the function $a(\eta)$
is no longer a classical solution of the background equations but a
quantum Bohmian trajectory of the quantized background, which may lead
to different power spectra.

\section{Spectrum of scalar perturbations in quantum cosmological models}
\label{sec:mixe}

Having obtained in the previous section the propagation equation for
the full quantum scalar modes, namely Eq.~(\ref{equacoes-mukhanov}),
in the Bohmian picture with the scale factor assuming the form
(\ref{at}), it is our goal now to solve this equation in
order to obtain the scalar perturbation power spectrum as predicted
by such models. In this section we obtain the analytical result for
long wavelength perturbations through a matching procedure, while in
the following we confirm our current findings by getting numerical
solutions.

We shall begin with the asymptotic behaviors. When $|T|\gg |T_0|$,
far from the bounce, one can write Eq.~(\ref{equacoes-mukhanov}) as

\begin{equation}
\label{Modes} v'' +\left[ \omega k^2
+\frac{2(3\omega-1)}{(1+3\omega)^2\eta^2}\right]\mu = 0,
\end{equation}
whose solution is
\begin{equation}
\label{Bessel} v = \sqrt{\eta} \left[ c_1(k) H^{(1)}_\nu
(\bar{k}\eta)+ c_2(k)
  H^{(2)}_\nu(\bar{k}\eta)\right],
\end{equation}
with
$$ \nu = \frac{3(1-\omega)}{2(3\omega+1)}, $$ $c_1$ and $c_2$ being
two constants depending on the wavelength, $H^{(1,2)}$ being Hankel
functions, and $\bar{k}\equiv \sqrt{\omega}k$.

This solution applies asymptotically, where one can impose vacuum
initial conditions as in \cite{MFB} \be v_\mathrm{ini} =
\frac{\ex^{i \bar{k} \eta}}{\sqrt{\bar{k}}},\label{v} \en which
implies that
$$ c_1=0 \quad \hbox{and} \quad c_2=\lP \sqrt{\frac{3\pi}{2}}
\ex^{-i\frac{\pi}{2} \left(\nu+\frac{1}{2}\right)}.
$$

The solution can also be expanded in powers of $k^2$ according to
the formal solution~\cite{MFB}
\begin{eqnarray}
\frac{v}{a} & \simeq & A_1(k)\biggl[1 - \omega k^2 \int^t \frac{\dd\bar
  \eta}{a^2\left(\bar \eta\right)} \int^{\bar{\eta}}
  a^2\left(\bar{\bar{\eta}}\right)\dd\bar{\bar{\eta}}\biggr]\nonumber
  \\ &+& A_2(k) \biggl[\int^\eta\frac{\dd\bar{\eta}}{a^2} - \omega k^2
  \int^\eta \frac{\dd\bar{\eta}}{a^2} \int^{\bar{\eta}} a^2
  \dd\bar{\bar{\eta}} \int^{\bar{\bar{\eta}}}
  \frac{\dd\bar{\bar{\bar{\eta}}}}{a^2} \biggr],\cr & & \label{solform}
\end{eqnarray}
up to order $ \mathcal{O}(k^{j\geq 4})$ terms. In Eq.~(\ref{solform}),
the coefficients $A_1$ and $A_2$ are two constants depending only on
the wavenumber $k$ through the initial conditions. Although this form
is particularly valid as long as $\omega k^2\ll a''/a$, \ie when the
mode is below its potential, Eq.~(\ref{solform}) should formally apply
for all times. In the matching region $\omega k^2\approx a''/a$, the
$\mathcal{O}(k^2)$ terms may give contributions to the amplitude, but
they do not alter the $k$-dependence of the power spectrum.

For the solution (\ref{at}), the leading order of the solution
(\ref{solform}) reads
\begin{eqnarray}
\frac{v}{a} &=& A_1 + A_2 T_0 a_0^{3(\omega-1)} \left(\arctan x +
\frac{\pi}{2}\right)\cr & \sim & A_1-A_2 T_0
a_0^{3(\omega-1)}\frac{1}{x},
\label{solmu0}
\end{eqnarray}
where $$x\equiv \frac{T}{T_0}.$$ In the last step we have taken the limit
$x\to-\infty$, and the constant $\pi/2$ was introduced in order for
$A_1$ represent the constant mode when it enters the potential.

Propagating this solution to the other side of the bounce, in the
expanding epoch, \ie the limit for $x\to +\infty$, yields
\begin{eqnarray}
\frac{v}{a} &\sim& A_1 + \biggl(\pi-\frac{1}{x}\biggr)
a_0^{3(\omega-1)}T_0 A_2 \cr &=&\left(A_1 + \pi a_0^{3(\omega-1)} T_0
A_2\right)+\frac{1}{x} a_0^{3(\omega-1)}T_0 A_2 .\label{specEND}
\end{eqnarray}
Note that there is a mixing in the constant part of the mode when it
passes through the bounce. Hence, in these types of bouncing models,
the bounce has important consequences for the final power spectrum.

In order to find the $k$-dependence of $A_1$ and $A_2$, we match $v$
and $v'$ in Eqs.~(\ref{solmu0}) and (\ref{Bessel}), obtaining, to
leading order,
\begin{eqnarray}
A_1 &=& \left[ \frac{3\omega-1}{3\alpha\left(\omega-1\right)} \tilde
C + a_0^{3\omega-1} T_0 \beta \tilde D \right]
\biggl(\frac{\bar{k}}{k_0}\biggr)^\frac{3\left(1-\omega\right)}
{2\left(3\omega+1\right)},\label{A1}\\ A_2 &=& -\left[
\frac{2a_0^{1+3\omega}}{3\beta\left(1-\omega\right)T_0} \tilde C -
\alpha \tilde D \right] \biggl(\frac{\bar{k}}{k_0}
\biggr)^\frac{3\left(\omega-1\right)}{2\left(3\omega+1\right)},\label{A2}
\end{eqnarray}
from which stem the spectral behaviors
\begin{equation}
A_1 \propto
k^{\frac{3\left(1-\omega\right)}{2\left(3\omega+1\right)}},\ \ \ \
A_2 \propto k^{\frac{3\left(\omega-1\right)}{2\left(3\omega+1\right)}},
\end{equation}
with
$$ \alpha = \left[
\frac{9\left(1-\omega\right)^2}{2\left|1-3\omega\right|}
\right]^{-1/(1+3\omega)}, \ \ \ \ \beta = \left[
\frac{9\left(1-\omega\right)^2}{2\left|1-3\omega\right|}
\right]^{\frac{1-3\omega}{2(1+3\omega)}},
$$
\begin{equation}
\tilde C = \sqrt{T_0} a_0^{(3\omega-1)/2}
c_2\sqrt{k\eta_{_\mathrm{M}}} H_\nu^{(2)}(k\eta_{_\mathrm{M}}),
\end{equation}
and
\begin{eqnarray} \tilde D &=&
  a_0^{(1-3\omega)/2} T_0^{-1/2} \frac{c_2}{2} \Bigg\{
\frac{H_\nu^{(2)}(\bar{k}
\eta_{_\mathrm{M}})}{\sqrt{\bar{k}\eta_{_\mathrm{M}}}} \cr & &
\hskip3mm + \sqrt{\bar{k}\eta_{_\mathrm{M}}} \left[
H_{\nu-1}^{(2)}(\bar{k}\eta_{_\mathrm{M}}) -
H_{\nu+1}^{(2)}(\bar{k}\eta_{_\mathrm{M}})\right] \bigg\},
\end{eqnarray}
with
$$k\eta_{_\mathrm{M}}=\frac{\sqrt{2|1-3\omega|}}{1+3\omega}$$ and
\be
k^{-1}_0=T_0a_0^{3\omega-1}. \label{k0}
\en

The coefficients $A_1$ and $A_2$ each contain a power-law behavior in
$k$. Because $0<\omega<1$, the power in $A_2$ [Eq.~(\ref{A2})] is
negative definite and that in $A_1$ [Eq.~(\ref{A1})] is positive
definite. Therefore, $A_2$ is the dominant mode, while $A_1$ provides
the sub-dominant mode.

The relation between the Bardeen potential $\Phi$ and $v$ is given by
Eq.~(\ref{vinculo-simples}). Using Eqs.~(\ref{solform}), (\ref{at})
and (\ref{jauge}) in order to change variables to $x$,
Eq.~(\ref{vinculo-simples}) leads to
\begin{eqnarray}\label{solform2}
& &\Phi a^{(1+3\omega)/2}\propto \biggl(\frac{v}{a}\biggr)' = -
\omega k^2
  \frac{{A}_1(k)}{a^2}
  \int^{x} T_0
  a^2\left(\bar{x}\right)\dd\bar{x}\nonumber
  \\ &+& \frac{{A}_2(k)}{a^2}\biggl[1 - \omega k^2
  \int^x \dd\bar{x}a^2 \int^{\bar{x}}
  \frac{T_0^2}{a_0^2(1+{\bar{\bar{x}}})^2}
  \dd\bar{\bar{x}}\biggr] + \mathcal{O}(k^4)\nonumber\\
  &\simeq&- \omega k^2
  \frac{{A}_1(k)}{a^2}
  \int^{x} T_0
  a^2\left(\bar{x}\right)\dd\bar{\eta}\nonumber
  \\ &+& \frac{{A}_2(k)}{a^2}\biggl[1 - \omega k^2\frac{T_0^2}{a_0^2}
  \int^x \dd\bar{x}a^2 (\arctan\bar{x}+\pi/2)\biggr],\nonumber\\
\end{eqnarray}
where the constant $\pi/2$ was introduced when performing the last
integral in order for ${A}_1$ to represent the constant mode at
large negative values of $x$ when $v_k$ is entering the potential,
as before. At large positive values of $x$, when $v_k$ is leaving
the potential and $x\propto\eta^{3(1-\omega)/(1+3\omega)}$, the
constant mode of $\Phi$, like $v$, mixes $A_1$ with $A_2$. In this
region, taking into account that $A_2$ dominates over $A_1$, we
obtain:
\begin{equation}
\Phi \propto
k^\frac{3\left(\omega-1\right)}{2\left(3\omega+1\right)}
\biggl[\mathrm{const.}+\frac{1}{\eta^{(5+3\omega)/(1+3\omega)}}\biggr].
\end{equation}
The power spectrum
\begin{equation}
\label{PS} \mathcal{P}_\Phi \equiv \frac{2 k^3}{\pi^2}
\left| \Phi \right|^2,
\end{equation}
then reads
\begin{equation}
\mathcal{P}_\Phi \propto k^{n_{_\mathrm{S}}-1},
\label{powspec}
\end{equation}
and we get
\begin{equation}
\label{indexS} n_{_\mathrm{S}} = 1+\frac{12\omega}{1+3\omega}.
\end{equation}

For gravitational waves (see Ref.~\cite{tens2} for details), the
equation for the modes $\mu = h/a$ reads
\begin{equation}
\label{mu} \mu''+\left( k^2 +2\Ka -\frac{a''}{a} \right)\mu =0,
\end{equation}
yielding the power spectrum for long wavelengths:
\begin{equation}
\label{PT} \mathcal{P}_h \equiv \frac{2 k^3}{\pi^2}
\left| \frac{\mu}{a} \right|^2.
\end{equation}
In Ref.~\cite{tens2}, we have obtained
\begin{equation}
\mathcal{P}_h \propto k^{n_{_\mathrm{T}}},
\label{powspect}
\end{equation}
with, as for the scalar modes,
\begin{equation}
\label{indexT} n_{_\mathrm{T}} = \frac{12\omega}{1+3\omega}.
\end{equation}

Note that in the limit $\omega\rightarrow 0$ (dust) we obtain a scale
invariant spectrum for both tensor and scalar perturbations.  This
result will be confirmed in the next section through numerical
calculations which will also give the amplitudes.  However, it is not
necessary that the fluid that dominates during the bounce be dust. The
dependencies on $k$ of $A_1$ and $A_2$ are obtained far from the
bounce, and they should not change in a transition, say, from matter
to radiation domination in the contraction phase, or during the
bounce. The effect of the bounce is essentially to mix these two
coefficients in order for the constant mode to acquire the scale
invariant piece. Hence, the bounce may be dominated by another fluid,
like radiation. If while entering the potential the fluid that
dominates is dust-like, then the spectrum should be almost scale
invariant. Note also that since we assume an ordinary matter fluid,
the equation of state is positive definite, being, in the most
pressure-less case, obtained as a result of the quantum non vanishing
mean square velocity. Thus, we do expect a blue spectrum.

\begin{figure}[t]
\vskip5mm\includegraphics[width=8.5cm]{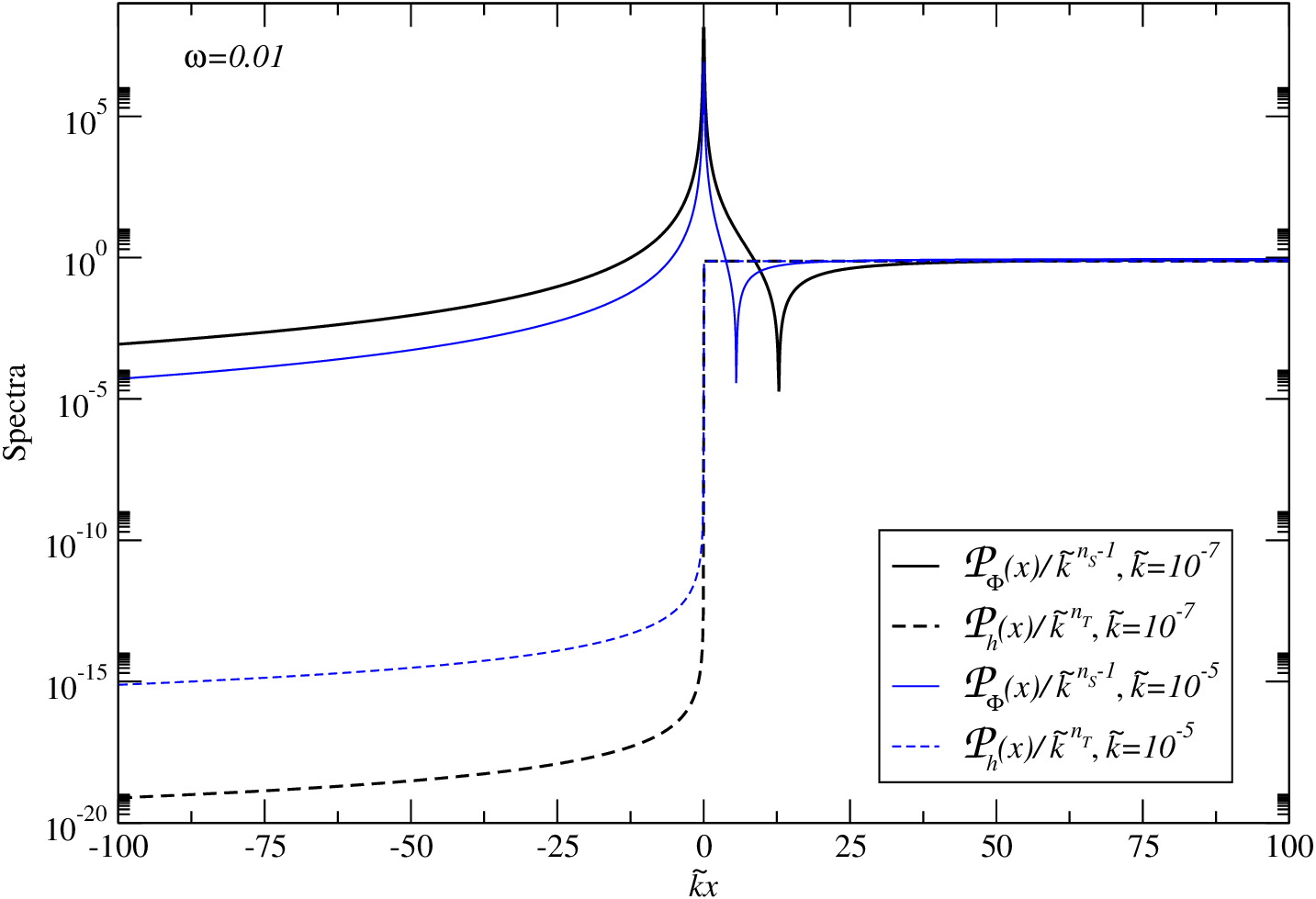} \caption{Rescaled
power spectra for scalar and tensor perturbations as functions of time
for $\omega=0.01$ and two different values of $\tilde k$. It is clear
from the figure that not only both spectra reach a constant mode, but
also that this mode does behave as indicated in Eqs.~(\ref{indexS})
and (\ref{indexT}). It is purely incidental that the actual constant
value of both modes are very close for that particular value of
$\omega$. The constant values obtained in this figure are the one used
to derive the spectrum below. In this figure and the following, the
value of $n_{_\mathrm{S}}$ used to rescale $\Phi$ is the one derived
in Eq.~(\ref{indexS}), thus proving the validity of the analytic
calculation.} \label{fig:v02}
\end{figure}

\section{Numerical results}\label{sec:nume}

In this section we will confirm the spectral indices of scalar
perturbations obtained above, and obtain the amplitude of the
scale invariant mode. The dynamical mode equation is expressible
in terms of the function $S \equiv
a^{\frac{1}{2}(1-3\omega)}v/\sqrt{T_0}$ (the constant $\sqrt{T_0}$
being introduced in order for $S$ to be dimensionless),
namely~\cite{tens2} \be \ddot S + \left[ \tilde k^2
\left(1+x^2\right)^\frac{2(3\omega-1)}{3(1-\omega)}
-\frac{1}{\left(1+x^2\right)^2} \right] S =0, \label{eqS} \en
with, in this latter case, $x\equiv T/T_0$ and $\tilde k \equiv
c_\mathrm{s}kT_0/a_0^{1-3\omega}$. We can apply the vacuum initial
conditions \be v_\mathrm{ini} = \frac{\ex^{ic_\mathrm{s}
k\eta}}{\sqrt{c_\mathrm{s} k}}, \label{v1ini} \en with the sound
velocity $c_\mathrm{s} = \sqrt{\omega}$ a constant. It is clear
here that one must insist upon not having $\omega=0$ in order to
be able to put these initial conditions. Again, anyway, the sound
velocity, even in a matter dominated phase, is not expected to be
vanishing identically.

\begin{figure}[t]
\vskip5mm\includegraphics[width=8.5cm]{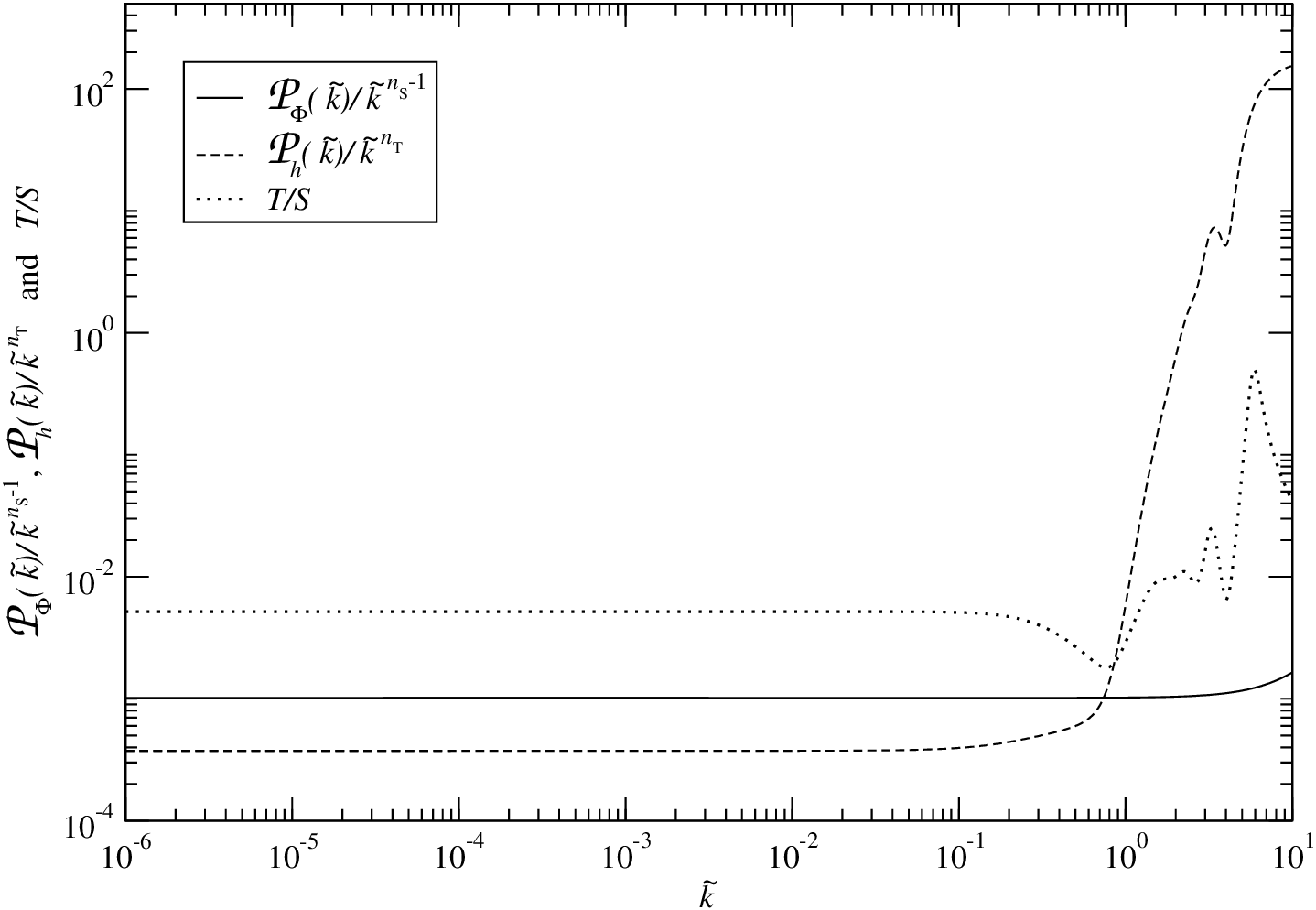} \caption{Rescaled
power spectra for $\omega=8\times 10^{-4}$, corresponding to the
conservative maximum bound on the deviation from a scale invariant
spectrum $n_{_\mathrm{S}} =1.01$, as function of $k$. The scalar
spectrum $\mathcal{P}_{\Phi}$ is the full line, while the dashed line
is the gravitational wave spectrum $\mathcal{P}_h$. Also shown is the
ratio $T/S$ (dotted); in this case, the $T/S\simeq 5.2\times 10^{-3}$,
i.e. almost two orders of magnitude below the current limit. This case
has a typical bounce length-scale of $L_0\sim 1.47 \times 10^3\lP$. The
amplitude of the modes is obtained as the constant part of
Fig.~\ref{fig:v02}.}
\label{fig:v03}
\end{figure}

Defining \be L_0 = T_0 a_0^{3\omega}, \label{L0} \en the curvature
scale at the bounce (the characteristic bounce length-scale), namely,
$L_0\propto 1/\sqrt{R_0}$ where $R_0$ is the scalar curvature at the
bounce, from which one can write \be \tilde{k}=\frac{k}{a_0}L_0\equiv
\frac{L_0}{\lambda_{\mathrm {phys}}^{\mathrm {bounce}}},
\label{tildek} \en one obtains the scalar perturbation density
spectrum as a function of time through \be \mathcal{P}_\Phi =
\frac{2(\omega+1)}{\pi^2 \sqrt{\omega} \left(1-\omega\right))^2 \tilde
k} |f(x)|^2 \left(\frac{\lP}{L_0}\right)^2, \en where
$$
f(x) \equiv
\left(1+x^2\right)^{-\frac{1+3\omega}{3(1-\omega)}}\frac{\dd S}{\dd
  x} - x \left(1+x^2\right)^{-\frac{4}{3(1-\omega)}} S,
$$ while the tensor power spectrum is \be \mathcal{P}_h =
\frac{2\tilde k^3}{\pi^2} \frac{|\bar v|^2}{1+x^2}
\left(\frac{\lP}{L_0}\right)^2,\en in which the rescaled function
$\bar v$, defined through~\cite{tens2}
$$
\bar v \equiv \frac{a^{\frac{1}{2}(1-3\omega)} \mu}{\lP\sqrt{T_0}},
$$ satisfies the same dynamical equation (\ref{eqS}) with
$c_\mathrm{s}\to 1$, with $\mu$ subject to initial condition \be
\mu_\mathrm{ini} = \sqrt{\frac{3}{k}} \lP \ex^{-ik\eta}.
\label{muini}
\en

{}From the above defined spectra, one reads the amplitudes \be
A_{_\mathrm{S}}^2 \equiv \frac{4}{25} \mathcal{P}_\zeta \label{AS2}\en
and \be A_{_\mathrm{T}}^2 \equiv \frac{1}{100} \mathcal{P}_h,
\label{AT2}\en where we assume the classical relation between $\Phi$
and the curvature perturbation $\zeta$ through \be \zeta =
\frac{5+3\omega}{3(1+\omega)}\Phi \ \ \to \ \ \mathcal{P}_\zeta =
\left[\frac{5+3\omega}{3(1+\omega)}\right]^2
\mathcal{P}_\Phi,\label{zetaPhi} \en to obtain the observed spectrum.
Since both spectra are identical power laws, and indeed almost scale
invariant power laws, the tensor-to-scalar ($T/S$) ratio, defined by
the CMB multipoles $C_\ell$ at $\ell=10$ as \be \frac{T}{S} \equiv
\frac{C_{10}^{\mathrm{(T)}}}{C_{10}^{\mathrm{(S)}}} =
\mathcal{F}(\Omega, \cdots )
\frac{A_{_\mathrm{T}}^2}{A_{_\mathrm{S}}^2}, \label{TsurS}\en can
easily be computed (see, e.g., \cite{MS00} and references therein). In
Eq.~(\ref{TsurS}), the function $\mathcal{F}$ depends entirely on the
background quantities such as the total energy density relative to the
critical density, i.e. $\Omega$, among others. It does not depend on
primordial physics parameters, supposed to be included in the
amplitudes of the spectra; in other words, it propagates the predicted
primordial spectra through the different epochs (radiation, matter and
cosmological constant dominated) whose characteristics are fixed by
observational data. For Eq.~(\ref{TsurS}) to hold true, both scalar
and tensor spectra ought to be power laws, as they are in our model.

Eq.~(\ref{TsurS}) permits to compare primordial cosmological effects
such as the ones derived here with current observations. This means
that our model must somehow be connected with the observed universe,
which we take to be the so-called concordance model (the one having a
cosmological constant accounting for $\Omega_\Lambda\simeq 0.7$ of the
total density). The relevant value of the function $\mathcal{F}$ can
be obtained from the slow-roll result: in this case, on
finds~\cite{MS00} $A_{_\mathrm{T}}^2/A_{_\mathrm{S}}^2 \sim 16
\epsilon$, with $\epsilon$ the usual slow-roll parameter, whereas the
consistency equation demands that
$C_{10}^{\mathrm{(T)}}/C_{10}^{\mathrm{(S)}} \simeq 10\epsilon$. For
the concordance model, one thus has $\mathcal{F} \sim 10/16 \simeq
0.62$. This is the value we shall use to compare our models with
observations.

We have solved these equations numerically and obtained the values
of the free parameters which best fit the data. First, we assume a
spectral index limited by $n_{_\mathrm{S}} \lesssim 1.01$, an
admittedly conservative constraint, which, given Eq.~(\ref{indexS}),
provide the already severe bound on the equation of state
$\omega\lesssim 8\times 10^{-4}$. Constraining $A_{_\mathrm{S}}^2 =
2.08\times 10^{-10}$ then implies the characteristic bounce
length-scale $L_0$ to be
$$
L_0 \gtrsim 1500 \lP,
$$
a value consistent with our use of quantum cosmology: it is indeed
in the kind of distance scale ranges that one expect quantum effects to be
of some relevance, while at the same time the Wheeler-de Witt equation
to be valid without being possibly spoiled by some discrete nature of
geometry coming from loop quantum gravity, and/or stringy effects.
The tensor-to-scalar ratio is, in this case, much
lower than the current WMAP3 constraint~\cite{contrainte} $T/S<0.21$
as we obtained $T/S\simeq 5.2\times 10^{-3}$. It is interesting to
note that when the $T/S$ constraint is almost reached, this gives a
numerical estimate for $\omega$ which is $\omega\simeq 8.5\times
10^{-2}$, and hence a spectral index $n_{_\mathrm{S}}\simeq 1.81$,
i.e. much larger than the current constraint. In this case also, one
finds $L_0\simeq 350\lP$. Hence, it is not necessary to explore these
possibilities.

Let us briefly discuss the validity of the model during the bounce
epoch. It is indeed not enough that the Wheeler-de Witt equation
applies in the regime one considers, but we should also make clear
that the backreaction is not going to dominate at the bounce. That
would for instance be the case if the Bardeen potential grows large
enough. In other words, we must ensure that $|\Phi(x)|\ll 1$ at all
times, a condition which, in view of Fig.~\ref{fig:v02}, might not be
easily satisfied. Indeed, in a way almost independent of $\omega$, one
finds that the spectrum $\mathcal{P}_\Phi$, once normalized to the CMB
for late times, is of order unity (depending only very little on the
values of $k$ and $\omega$) at its maximum, close to the bounce. This
is not so much of a problem since what needs to be small is the
Bardeen potential itself, and its Fourier mode reads
\begin{equation}
\Phi_k = \frac{f(x)}{\tilde k^2}
\frac{\sqrt{\omega(\omega+1)}}{1-\omega} \left(\frac{\lP
L_0^{1/2}}{a_0^{3/2}}\right),
\end{equation}
having an explicit dependence in $a_0$. As this latter parameter is
not constrained (in fact, it must be such that large physical scales
now must be much greater than $L_0$ at the bounce in order for our
approximation $\tilde{k}\ll 1$ be valid all along, see
Eq.~(\ref{tildek})), one can assume it is sufficiently large to yield
$\Phi_k\ll 1$ at all time and for all $\tilde k$ of cosmological
relevance. Note that $a_0$ could in principle be fixed by the
normalization of the scale factor at the bounce in terms of, say, the
Hubble constant today, to fix the scales $\lambda=a/k$ to be in the
observable range now. This will be done when we examine in a future
publication a more elaborated model containing not only dust but also
radiation, as discussed in the end of Sec.~III, in such a way that it
can be connected to a radiation dominated expansion phase before
nucleosynthesis.

\section{Conclusions}

Using quantum cosmology and the Bohm interpretation, thanks to which
one can define trajectories and a scale factor evolution with time, we
have obtained a simple model whose scalar and tensor perturbations can
be made arbitrary close to scale invariance. The model consists of a
classically contracting single dust perfect fluid in which the big
crunch is avoided through quantum effects in the geometry described by
the Wheeler-DeWitt equation, turning the Universe evolution to an
expanding phase, which soon becomes classical again.  This transition
is smoothly described by a Bohmian quantum trajectory containing a
bounce. Perturbations begin in a vacuum state during the contraction
epoch, when the Universe was very large and almost flat, and are
subsequently evolved in a fully quantum way through its all
history. Hence, we have presented a nonsingular model without horizons
where perturbations of quantum mechanical origin can be described all
along and generate structures in the Universe.

In order to match the CMB normalization, we find that the
characteristic length at the bounce must be of the order of a few
thousands Planck lengths, thereby making the model fully
consistent. Indeed, we should like to emphasize that one would expect
precisely the Wheeler-de Witt equation to be valid and important in
this regime (a sufficiently low scale for quantum gravity effects be
important, but not so low in order for not being affected by string
and/or loop quantum gravity effects).  Moreover, it predicts a
slightly blue spectrum, which may be seen as a drawback of the model,
although this point still deserves further experimental clarification,
say by the Planck mission. However, in more realistic and elaborated
models, other fluids must be considered, like radiation and dark
energy. It seems that adding these fluids will not spoil our results
as long as only a single fluid dominates at the bounce, and that dust
dominates when the scales of cosmological interest become greater than
the curvature scale in the classical contracting phase. If in that
period dark energy has some effect imposing a slightly negative
effective $w$, then one could even obtain a slightly red spectrum
instead of the blue one derived here. Besides, ironically coming back
to one of the original motivation for inflation~\cite{ebg}, we remark
on Fig.~\ref{fig:v03} that the gravitational potential shows a net
increase for large values of $\tilde k$ after the bounce has taken
place. That could initiate non-linear growth of the primordial
perturbations at very small scales (provided the values of $T_0$ and
$a_0$ are chosen conveniently). In turn, these would lead to the
formation of primordial black hole whose decay could initiate a
radiation dominated phase, as needed. These elaborations should also
consider other classical cosmological puzzles, namely the flatness and
remnants (e.g.  monopoles) problems, and baryogenesis. These issues
will be considered in future works.

Finally, we have obtained that once the constraint on the scalar index
is taken into account, the tensor-to-scalar ratio which follows is
predicted to be small. Thus, this category of model is falsifiable.

\section{Acknowledgments}

We would like to thank CNPq of Brazil for financial support. We would
also like to thank both the Institut d'Astrophysique de Paris and the
Centro Brasileiro de Pesquisas F\'{\i}sicas, where this work was done,
for warm hospitality. We very gratefully acknowledge various
enlightening conversations with J\'er\^ome Martin, Slava Mukhanov and
Luis Abramo. We also would like to thank CAPES (Brazil) and COFECUB
(France) for partial financial support.


\begin{thebibliography}{xx}

\bibitem{MFB} V.~F.~Mukhanov, H.~A.~Feldman, and R.~H.~Brandenberger,
Phys. Rep. {\bf 215}, 203 (1992).

\bibitem{cobe} G.~F.~Smoot \etal, Ap. J. {\bf 396}, L1 (1992);
  E.~L.~Wright \etal, Ap. K. {\bf 396}, L13 (1992).

\bibitem{WMAP} D.~N.~Spergel \etal, Ap. J. Suppl. {\bf 148}, 175
  (2003); D.~N.~Spergel \etal, {\tt astro-ph/0603449}, Ap. J. (2006).

\bibitem{falsify} G.~Veneziano, Phys. Lett. B {\bf 265}, 287 (1991);
M.~Gasperini and G.~Veneziano, Astropart. Phys. {\bf 1}, 317 (1993);
See also J.~E.~Lidsey, D.~Wands, and E.~J.~Copeland, Phys. Rep. {\bf
337}, 343 (2000) and G.~Veneziano, in {\sl The primordial Universe},
Les Houches, session LXXI, edited by P.~Bin\'etruy {\it et al.}, (EDP
Science \& Springer, Paris, 2000); J.~Khoury, B.~A.~Ovrut,
P.~J.~Steinhardt, and N.~Turok, \prd {\bf 64}, 123522 (2001); {\tt
hep-th/0105212}; R.~Y.~Donagi, J.~Khoury, B.~A.~Ovrut,
P.~J.~Steinhardt, and N.~Turok, JHEP {\bf 0111}, 041 (2001).

\bibitem{inflation} A.~Guth, \prd {\bf 23}, 347 (1981); A.~Linde,
Phys.  Lett. B {\bf 108},389 (1982); A.~Albrecht and P.~J.~Steinhardt,
\prl {\bf 48}, 1220 (1982); A.~Linde, Phys. Lett. B {\bf 129}, 177
(1983); A.~A.~Starobinsky, Pis'ma Zh. Eksp.  Teor. Fiz. {\bf 30}, 719
(1979) [JETP Lett. {\bf 30}, 682 (1979)]; V.~Mukhanov and G.~Chibisov,
JETP Lett. {\bf 33}, 532 (1981); S.~Hawking, Phys. Lett. B {\bf 115},
295 (1982); A.~A.~Starobinsky, Phys. Lett. B {\bf 117}, 175 (1982);
J.~M.~Bardeen, P.~J.~Steinhardt, and M.~S.~Turner, \prd {\bf 28}, 679
(1983); A.~Guth, S.~Y.~Pi, \prl {\bf 49}, 1110 (1982).

\bibitem{pinto} J.~Acacio de Barros, N.~Pinto-Neto, and
M.~A.~Sagioro-Leal, Phys. Lett. A {\bf 241}, 229 (1998).

\bibitem{pinto2} R.~Colistete Jr., J.~C.~Fabris, and N.~Pinto-Neto,
\prd {\bf 62}, 083507 (2000).

\bibitem{fabris}  F.G. Alvarenga, J.C. Fabris, N.A. Lemos and G.A. Monerat,
Gen.Rel.Grav. {\bf 34}, 651 (2002).

\bibitem{bounce} G. Murphy, Phys. Rev. {\bf D8}, 4231 (1973);
A. A. Starobinsky, Sov. Astron. Lett. 4, {\bf 82} (1978); M. Novello
and J. M. Salim, Phys. Rev. {\bf D20}, 377 (1979); V. Melnikov and
S. Orlov, Phys. Lett {\bf A70}, 263 (1979); P.~Peter and
N.~Pinto-Neto, \prd {\bf 65}, 023513 (2002); P.~Peter and
N.~Pinto-Neto, \prd {\bf 66}, 063509 (2002); V.A. De Lorenci,
R. Klippert, M. Novello and J.M. Salim, \prd {\bf 65}, 063501 (2002);
J. C. Fabris, R. G. Furtado, P. Peter and N. Pinto-Neto \prd {\bf 67},
124003 (2003); P. Peter, N. Pinto-Neto and D. A. Gonzalez, JCAP {\bf
0312}, 003 (2003).

\bibitem{tens1} P.~Peter, E.~Pinho, and N.~Pinto-Neto, JCAP {\bf 07},
  014 (2005).

\bibitem{tens2} P.~Peter, E.~Pinho, and N.~Pinto-Neto, \prd {\bf 73},
  104017 (2006).

\bibitem{scalar} E.~J.~C.~Pinho, N.~Pinto-Neto, {\it Scalar and Vector
  Perturbations in Quantum Cosmological Backgrounds}, {\tt
  hep-th/0610192}. 

\bibitem{halli} J.~J.~Halliwell and S.~W.~Hawking, \prd {\bf 31}, 1777
(1985).

\bibitem{hol} See for instance \eg P.~Holland, {\it The Quantum Theory
of Motion}, Cambridge University Press (Cambridge, UK, 1993) and
references therein. See also Ref.~\cite{pinto} for a cosmological
setting.

\bibitem{Schutz} B.~F.~Schutz, Jr., \prd {\bf 2} 2762 (1970); \prd
{\bf 4}, 3559 (1971).

\bibitem{MS00} J.~Martin and D.~J.~Schwarz, \prd {\bf 62}, 103520 (2000).

\bibitem{contrainte} J.~Martin and C.~Ringeval, JCAP {\bf 08}, 009
  (2006).

\bibitem{ebg} R.~Brout, F.~Englert, and E.~Gunzig, Ann. Phys. {\bf
  115}, 78 (1978).

\end{thebibliography}
\end{document}